# Paul Alsberg (1883-1965) et le transfert adaptatif du biologique au technique : un précurseur de la "cultural niche construction" ?

Paul Alsberg (1883-1965) and the adaptive transfer from biologic to technology: a precursor of "cultural niche construction"?

Gérald FOURNIER[1]

**RÉSUMÉ** – Nous proposerons dans cet article une présentation ainsi qu'une discussion des thèses de Paul Alsberg (1883-1965) concernant le supposé principe spécifique de l'évolution humaine. L'auteur maintient une différence de nature entre l'Homme et l'Animal en se fondant sur une opposition entre adaptation *somatique* – celle de l'animal – et adaptation *exosomatique* – celle de l'humanité ; les moyens d'adaptation se trouvant désormais à l'extérieur de l'organisme, par l'utilisation d'outils. Cette différence là n'est pas une simple différence d'*état*, mais bien de *dynamique* évolutive. L'homme n'est pas simplement ici « *Homo faber* », comme chez Bergson, mais produit et rendu possible par la technique ; une technique qui apparaît alors comme un véritable facteur d'hominisation. Ainsi, son principe d'*émancipation somatique* par utilisation d'outils s'inscrit, rétrospectivement, dans une logique de modification des pressions de sélection qui n'est pas sans rappeler la théorie contemporaine et féconde de la "*construction de niche*" (F. John Odling-Smee). Il semble dès lors possible de faire de Paul Alsberg, par les thèses qu'il énonça dès 1922, un des précurseurs importants de la "*cultural niche construction*".

**MOTS CLÉS :** TRANSFERT ADAPTATIF – ADAPTATION SOMATIQUE ET EXOSOMATIQUE – ÉMANCIPATION SOMATIQUE / CONTRAINTE SOMATIQUE – CONSTRUCTION DE NICHE – TECHNIQUE – HOMINISATION

**ABSTRACT** – We propose, in this paper, both a presentation and a discussion of Paul Alsberg's thesis on the supposed specificity principle of human evolution. The author maintains a difference of nature between Man and Animal relying on an opposition between "body-adaptation" – that of the Animal – and "extrabodily-adaptation" – that of Man in which the means of adaptation are switched outside of the organisms by tool-using. This difference is not a mere difference of *state*, but of evolutionary *dynamics*. Here, Man is not simply "*Homo faber*", as in Bergson's view, but produced and made possible by technique; a technique which then appears as an hominisation factor. Thus, his "principle of body-liberation" by tool-using is to be retrospectively understood as a part of the logics of the modification of selection pressure logics, which reminds us the seminal contemporary *niche construction* theory (F. John Odling-Smee). It seems therefore possible to make Paul Alsberg, from his 1922 work, one of the most important precursors of the *cultural niche construction* theory.

---

[1] Doctorant au *Laboratoire d'Études du Phénomène Scientifique*, (LEPS), E.A. 4148, *Université de Lyon*, *Université Lyon 1*.

# Introduction

> "The principle of animal evolution is that of compulsory adaptation to environment by means of the body: *the principle of body-compulsion*. The principle of human evolution is that of freeing Man from the compulsion of body-adaptation by means of artificial tools: *the principle of body-liberation*."[2]

L'objectif de cet article est de faciliter l'accès à la pensée d'un médecin du XXème siècle traitant de la différence anthropologique dans le cadre de la biologie évolutive. Cet intérêt pour Alsberg, nous le devons à un passage de Peter Sloterdijk dans *La domestication de l'être*[i]. D'origine allemande, il publie en 1922 *Das Menschheitsraetsel*, que l'on pourrait traduire par *L'énigme de l'humanité*. Ouvrage qui lui tenait à cœur, et brulé sous le régime d'Hitler, il en écrira lui-même une version anglaise peu de temps avant sa mort. C'est, faute de mieux, sur cette dernière version que nous nous appuierons. Il sera ici proposer une restitution de sa pensée, ainsi que quelques commentaires sur les forces et les faiblesses de sa théorie de l'homme et de l'adaptation, en résonnance avec des théories plus contemporaines. Le centre de sa théorie, sa ligne directrice, se constitue autour d'un clivage entre une évolution régie par le principe de *body-compulsion* et un autre principe, typiquement humain, de *body-liberation*. Corrélativement on doit ajouter les principes de *body-adaptation* et d'*extra-body adaptation*.

*Choix de traduction*

Afin de témoigner au mieux de la pensée de l'auteur, on traduira le principe de *body-compulsion,* non pas par *le principe corps-contrainte ou* par le *principe de contrainte corporelle*, mais par *principe de contrainte somatique*, avec son *adaptation somatique*[3] (*body-adaptation*) plutôt que corporelle. Quant aux concepts relatifs à la dynamique évolutive humaine on parlera du principe d'*émancipation somatique* (*body-liberation*) – au lieu de *suspension des corps*, ou *libération corporelle* – caractérisé par une *adaptation exosomatique*[4] (*extra-bodily adaptation*) plutôt qu'extracorporelle. Émancipation qui s'accompagne et a pour effet indirect un *délestage somatique* (*body-elimination*) ; une régression des anciennes caractéristiques somatiques, engendrant alors une "redistribution morphologique". Ce principe d'émancipation somatique veut englober deux choses : *Primo*, le fait que la tâche de l'adaptation ne repose plus sur les organismes, mais sur la technique ; *deuxio*, l'influence de ce nouveau type d'adaptation sur la dynamique évolutive de l'espèce en question, caractérisée par un délestage somatique ; par un *feedback* jouant sur son évolution ultérieure. En somme, quelque chose de l'ordre de *la construction de niche*[5] (Odling-Smee, 1988, 2003). Cette liberté dans la traduction, je l'espère, témoignera plus facilement du sens qu'a voulu donner l'auteur.

---

[2] Paul ALSBERG, *In Quest of Man*, *a biological approach to the problem of man's place in nature*, Pergamon Press, 1970, *chap*. V, p. 38.
[3] Le choix du *soma* ne suggère pas ici un renvoi à l'adaptation phénotypique et à son opposition au *germen*.
[4] Si le choix de ce terme nous a été suggéré ailleurs, on retrouve dans le *que sais-je ?* de Jean CHALINE, *L'évolution biologique humaine*, 1982, p. 118, le passage suivant : « H. Tintant (1975) qualifie cette évolution biologique d'endosomatique car elle se manifeste au niveau de l'organisme, de l'individu [ex : la tendance à la disparition des dents de sagesse, démontrant au passage que l'évolution humaine se poursuit]. Or, avec l'accès à la pensée réfléchie, l'Homme va introduire dans le monde un nouveau type d'évolution, pour laquelle le même auteur emploie le terme d'exosomatique. Exosomatique, cela signifie que l'Homme a la possibilité de s'adapter au milieu par des moyens situés en dehors de son organisme, par le truchement d'outils, et en se créant même un milieu artificiel ; donc sans que le corps ait besoin de se modifier. » J. Chaline nous donne la référence dans sa bibliographie : Henri TINTANT, « L'Homme, produit ou auteur de son évolution », *Mém. Acad. Sc., Arts et Belles-Lettres de Dijon*, t. 122, p. 1-12., 1975. Une telle vision se retrouve aussi (nous indique Jean-Michel TRUONG, 2001), dans K. R. C. POPPER & J. C. ECCLES, « L'évolution culturelle continue l'évolution génétique par d'autres moyens », *The Self and its Brain*, Routledge & Kegan Paul, 1984, p. 48.

Face à l'orthodoxie darwinienne privilégiant une *différence de degré*, la philosophie biologique de Paul Alsberg a ceci d'intéressant qu'elle continue à réserver à l'homme une *différence de nature*[ii] tout en étant particulièrement attentive à l'évolution humaine[6]. Cette différence spécifique est recouverte sous l'expression de « *body-liberation* » ou encore d'« *extra-bodily adaptation* ». Comme pour Bergson[iii], chez Alsberg, l'homme n'est homme qu'en tant qu'il est *Homo faber*[7]. Mais, plus précisément, il ne s'agit pas là uniquement de penser la spécificité humaine mais bien *sa spéciation*[8] : l'anthropogenèse ou hominisation.

## 1) De l'Adaptation Somatique à l'Adaptation Exosomatique

### 1.1) L'adaptation et la contrainte somatiques, "body-adaptation" et "body-compulsion"

Certes, il peut sembler étrange de concevoir l'adaptation, dans la perspective de la biologie évolutive, comme autre chose que somatique. Le "cognitif" ou le système nerveux ne sont rien d'autre que du somatique et la biologie n'a pas à souffrir du dualisme philosophique corps/esprit. L'adaptation somatique intervient, chez Alsberg, comme la désignation du schème classique d'évolution ; référentiel qui devra permettre de penser la spécificité, le monopole même, du schème d'évolution humaine. La position d'Alsberg est, à ce sujet, claire : "Man appears in his evolution to be a singular exception to the universal scheme of body-compulsion."[iv]. Il faut donc concevoir par « adaptation somatique », l'adaptation générale par sélection naturelle où les espèces sont façonnées par leurs interactions avec le milieu abiotique et biotique, par les pressions intraspécifiques et interspécifiques, au cours des différentes ères géologiques. Ainsi, pour l'auteur, l'adaptation animale se fait simplement par les moyens du corps, avec toute la diversité que nous constatons. Le principe d'adaptation somatique est un schème d'évolution caractérisé par le fait que le corps soit l'unique "ce sur quoi" pèse la tâche de l'adaptation. Alsberg nomme cette modalité première le « *principe de contrainte somatique* »[v].

---

[5] Niche construction: "The process whereby organisms, through their metabolism, their activities, and their choices, modify their own and/or each other's niche. Niche construction may result in changes in one or more natural selection pressures in the external environment of populations. Niche-constructing organisms may alter the natural selection pressures of their own population, of other populations, or of both." ODLING-SMEE, LALAND & FELDMAN, *Niche Construction, the neglected process in evolution*, Princeton, 2003, glossary, p. 419.

[6] "From the standpoint taken in this book, it was, conversely, Man who broke away from the central line, the animal line, and thus became Man, while the Apes continued to follow the old line, and thus remained Animals. (…) In conclusion, the Ape is neither a sort of degenerate or brutalized Man; nor is Man a sort of transformed Ape, for the simple reason that he is no longer an Ape. Phylogenetically, both hang together by the bonds of blood relationship; biologically, they are separated by the unbridgeable chasm of opposite evolutionary principles. From this point of view it follows that the Linnaean system of classification in which Man was placed with the Apes in the Order of the Primates requires urgent revision." P. ALSBERG, *In Quest of Man*, XVIII, p. 163.

[7] "For, from the critical point in time, when Man consistently and persistently turned out to the use of artificial tools in preference to, and actual abandonment of, the natural means of his body, and by so doing made tool-use the soul and destiny of his evolution, he was truly Man." (*Ibid.*, XVI, p. 138).

[8] "*Primeval Man may be defined as the one descendant of an anthropoid stock who integrated tool-use into his evolutionary mechanism*." (*Ibid.*, XIV, p. 117). Ou encore : "There is also some perplexity among paleontologists over the question of how long it may have taken Man to develop from his anthropoid ancestry into true Man. Our theory indicates that it must have been a rapid transition, since it was simply the change from the animal to the human principle, as implemented by the systematic use of artificial tools, that created Man." (*Ibid.*, XVIII, p. 155).

### 1.2) Du somatique à l'exosomatique

Sa nudité, sa polyvalence et sa technique font que l'homme puisse apparaître comme étranger au monde, trop différent qu'il est de l'animal. Cependant, comme tout biologiste de l'évolution, Alsberg pense que l'homme n'a pas pu devenir ainsi sans traverser une période animale ; que l'homme, tel qu'il est, avec cette apparente étrangeté, reste un produit évolutif qu'il s'agit de comprendre. Alsberg s'interroge et fixe le problème à résoudre :

> "But if he, like all Animals, was physically well adapted to his surroundings, then the present unnatural state of Man admits of only one conclusion: that those vital adaptations, still possessed by his animal precursor, must have been lost to him in the course of his evolution."[vi]

La question est donc de savoir quel phénomène a fait perdre au prototype pré-humain ses adaptations somatiques. À cet effet, Alsberg propose brièvement l'hypothèse d'une baisse locale et hasardeuse (isolement géographique) des pressions de prédation subies par le pré-humain, rendant alors inutiles ses moyens de protection naturels (dépérissement des anciennes formes). Ensuite, après cette phase de ré-adaptation à des conditions d'existence plus clémentes, (par délestage somatique) il aurait pu, suite au développement correspondant de son intelligence, développer les outils lui permettant de compenser les lacunes physiques (toutes somatiques) ainsi acquises[9].

## 2) L'Adaptation Exosomatique

### 2.1) "Extra-bodily adaptation" et "body-liberation"

Ce qu'Alsberg défend, quant à la trajectoire évolutive de l'humanité et l'hominisation, est ce que l'on pourrait appeler *un report ou transfert adaptatif* du somatique vers le technique, tout en prenant en compte l'influence de cette "nouveauté" adaptative sur sa morphologie actuelle. L'auteur

---

[9] "If this be so, the problem arises as to how to fit in his total loss of natural protective equipment with the exigencies of the universal struggle for existence. It has been suggested that Primeval Man may, for a while, have been exempted from that fierce struggle if by chance he lived in a region free from dangerous beasts, and here enjoyed a life of paradisial happiness with no need of defensive weapons; and that in consequence of this peaceful existence, his original means of protection would have gradually disintegrated, whilst his mental abilities would have correspondingly increased. When in the course of time the situation grew worse, and hardship and danger had again to be faced, his intellect had meanwhile developed to such a degree that it enables him to make up for his physical deficiencies by *the invention of artificial tools*." (*Ibid.*, V, pp. 31-32).

explique ce passage d'une adaptation somatique classique à une adaptation exosomatique en ces termes :

> "In the case of Man evolution appears to have taken a new direction in which adaptation to environment was no longer entrusted to the body but was implemented by *artificial tools*. In short, *bodily adaptation* switched over in the case of Man to *extra-bodily* adaptation**.** Consequently, with the duties of adaptation being transferred from the body to the artificial tool, human evolution took a course in which all effort that hitherto (in the animal ancestry) served the task of developing the highest possible grade of body-adaptation was henceforth (with Man) devoted to the highest possible development of the artificial tool. The final result on the one hand was the establishment of a gigantic realm of highly finished artificial tools and contrivances towering *around* Man, while on the other hand there was a singular deterioration of the body itself, which made his maintenance of life totally dependent upon the use of artificial tools."[vii]

C'est désormais, dans le cas de l'homme, *sur l'outil que pèse la tâche de l'adaptation,* ce qui permet à l'auteur de donner à ce dernier un statut à part, contre la logique même de la biologie évolutive. En cela réside le dépassement essentiel qui pousse notre auteur à défendre une différence de nature entre l'humanité et le reste du vivant. On notera aussi, dans cette dernière référence, le fait que l'homme devienne complètement dépendant de la technique, ce qui nous fait comprendre que la trajectoire évolutive de l'humanité pourrait être qualifiée de *"technosymbiotique"*[10]. Cette dernière appellation n'aurait probablement pas déplu à Paul Alsberg. Tel serait le rôle fondamental et ancien de la technique, comme véritable facteur d'hominisation, par *modification avantageuse de la structure des pressions de sélection.*

Cette perspective ici énoncée l'oppose à André Leroi-Gourhan et sa théorie de la technique comme *projection corporelle*[viii]. Une différence majeure est à l'œuvre, car, pour Alsberg, l'utilisation d'outils ne prolonge pas le corps, elle l'élimine :

> "The **"organ-projection"** theory has found so much favour because the organs, particularly hand and arm, are most often actively brought into play when using a tool – as, for example, in the act of swinging a hammer. Still more, the performance of the hammer is practically like that of an extended and reinforced hand. Yet, undeniably, the fact is that it is the hammer that drives the nail into the wall, not the hand; and that the hammer is even made for the special purpose of doing the work *instead of the hand*. Still the hammer cannot work unless it is put into operation, and it is only for this task of *operating* the hammer that the hand is engaged. As to the actual work done and intended to be done by the hammer, the hand is, in fact, *eliminated*."[ix]

C'est justement le fait d'utiliser l'outil qui a une fonction biologique et évolutive, ce qui, dans l'esprit d'Alsberg, constitue le fait même de ne plus utiliser le registre somatique naturel et premier pour s'adapter. Prendre pleinement conscience de ce fait, de cette redistribution anatomique et relative aux pressions de sélection, implique l'insatisfaction envers la simple théorie de la technique comme projection d'organe. Au contraire, soutient Alsberg, "Tool use thus became the dynamic principle of human evolution."[x] Cette adaptation, par des moyens désormais exosomatiques, fait que l'homme se rend de plus en plus indépendant de certaines contraintes environnementales, donc modifie la structure des pressions de sélection auparavant actives et fonctionnelles, engendrant alors une modification de son propre être et de sa trajectoire évolutive ultérieure : "Once the tool had come into the possession of Man, he was himself possessed by it, and he put into it all his soul and strength."[xi]

---

[10] Par humanité *"produite par technosymbiose"*, j'entends, produite par une longue coévolution avec la technique, d'abord élémentaire puis généralisée. Une coévolution pouvant rendre compte à la fois de sa spéciation, de ses caractéristiques morphologiques, et, dans une certaine mesure, de son avenir du fait de sa dépendance croissante (*obligate niche construction*) envers ce *"symbiote"* très particulier.

## 2.2) Le feedback évolutif

L'activité technique, dès lors qu'on l'insère dans une temporalité évolutive, déclenche une rétroactivité apte à faire changer la morphologie de l'espèce qui la mobilise. Ce changement, en raisonnant avec le pré-humain, doit faire entrer en régression (au même titre que le défaut d'usage, mais ici par une voie strictement conforme à la logique sélective non « lamarckienne ») les anciens équipements somatiques[11]. On comprend alors que la perte des équipements naturels et premiers de protection est le résultat et non la cause de l'activité technique :

> "From the standpoint of this book the very opposite assumption holds good: his loss of natural equipment was the result rather than the cause of his taking to tools. When we use a tool we bring the principle of body-liberation into play, no matter how small be the extent of body-liberation demanded. (…) In either case it is an artificial contrivance, a "tool", that we are using for adaptation to environment *instead of* effectuating adaptation by means of our body."[xii]

L'adaptation somatique classique, enfermant l'espèce dans ses relations proie-prédateur, dans une éternelle « course aux armements »[xiii], n'est plus, à partir de ce moment, le moteur exclusif et principal de l'adaptation et de la réussite de l'espèce humaine dans ses interactions biologiques. Tel est le nouveau schème d'évolution dans lequel l'*Homo* devenu *sapiens* eut la chance de se trouver. Ce nouveau schème le rend ainsi véritablement différent et à distance de l'Animal et de ses limitations :

> "The use of artificial tools does not simply mean replacing one means of adaptation, the body-organ, by another more efficient one, the extra-bodily tool. What it really means is that in human evolution *the task of adaptation to environment has switched over the body to the tool*. (…) From an evolutionary point of view it would appear that by renouncing his own bodily resources and choosing instead his extra-bodily scheme of adaptation *Man shook off the yoke of compulsory bodily adaptation* and thereby *freed* himself by means of his superior artificial tools from the natural restrictions and limitations inherent in the animal scheme of body-adaptation."[xiv]

Cette action sur l'environnement se retourne alors sur l'espèce agissante : "With Man, therefore, it is no longer the body which develops into specialized forms of adaptation, but it is the artificial tool on which evolution is henceforth focused, whereas the body only follows secondarily this trend of evolution and thereby forfeits its original adaptive outfit."[xv] Si cette perte d'équipements naturels, est le résultat de l'activité technique, il en va de même avec le couple technique/céphalisation. Pour Alsberg la technique ne serait donc pas le résultat, mais la cause de l'accroissement cérébral menant à l'*Homo sapiens*[12]:

> "A further instance of progressive growth is the enormous expansion and elaboration of the human brain. As will be discussed later, its development from the smaller and more primitive simian brain must again be supposed to have been greatly influenced by the use of artificial tools; and here also it would seem progress was brought at price: the decline of instinct as an element of former adaptive outfit."[xvi]

---

[11] "it was the elimination of the human body by means of artificial tools from the duty of adaptation that subsequently led to the loss of original adaptational equipment." ALSBERG, *op. cit.*, V, pp. 38-39.

[12] "Here in any case due account is taken of the intimate interplay between tool-use and brain development in a situation which must have taxed the mental faculties to the extreme. Although we may reasonably assume that brain growth, this outstanding feature of human evolution, had already occurred with the start of Man, it would not carry the implication that Man started from an already advanced brain condition, that is, from a brain that transcended greatly that of the intelligent Chimpanzee." (*Ibid.*, XVI, p. 136).

L'accroissement de taille du cerveau serait ainsi causé par l'utilisation d'outils, ce qui, de plus, semble être en adéquation avec les données de la paléoanthropologie[xvii] où la technique et la bipédie ont précédé cette augmentation rapide du volume cérébral[13]. C'est ensuite en modifiant les pressions de sélection pesant sur son espèce, par sa capacité à configurer, à modeler son environnement[xviii], que l'homme a pu devenir ce qu'il est aujourd'hui.

# 3) Discussion : forces et faiblesses de sa théorie

### 3.1) La position de l'homme : une différence de nature et non de degré[14]

Que l'homme soit partie intégrante de la nature n'empêche pas notre auteur de considérer ce dernier comme au-delà et au-dessus de l'animal[xix]. Cette différence de modalité adaptative ou de schème évolutif conduit Alsberg à soutenir une vision de l'homme assez proche de celle de H. Bergson[xx], comme « âme de la Terre »[xxi], comme animal libre physiquement et spirituellement :

> "From time immemorial Man has been led to believe in his "higher destination", not so much for his brilliant achievements in material things, his wonders of technology, but foremost for his profound supremacy of mind, as it is so impressively evidenced in his works of science, art, philosophy, and religion. The old dogma, therefore, not only lifted Man upon a "separate place in Nature", but also claimed for Man a place "outside and above Nature", in contradiction from the Animal that had its place "in and below Nature". (…) According to our theory, Man owes his unique existence, and all that evolution has brought him, to his exclusive principle of body-liberation which bestowed upon him the gift of *freedom*. Two spheres of freedom can be discerned: his *physical* freedom freeing him physically from the bondage of his body, and his *spiritual* freedom making him mentally independent of the strings of body-compulsion."[xxii]

Alsberg voit en effet l'homme comme le représentant d'un nouveau type de vie[15], et sinon comme objectif de la création, comme ce par quoi l'évolution travaille :

> "If it is right to say that Man not only forms the apex of the pyramid of living beings but is the very crown of the creation, it follows that evolution in its upward movement virtually *worked towards the rise of Man*."[xxiii]

---

[13] La bipédie semble avoir démarrée il y a 4 Ma. Quant à la technique, un problème important réside dans la distinction entre l'usage (*tool-users*) et la fabrication d'outils (*tool-makers*). Que faire, par exemple, de ce que Dart a nommé l'industrie ostéo-donto-kératique, parlant des Australopithèques comme des *tool-users* et non *makers* (Ruffié, *De la biologie à la culture*, vol. 1, p. 248). Or, dans la théorie ici développée, c'est l'usage qui compte. Les traces de *fabrication* d'outils tournent autour de 2.5 Ma. En ce qui concerne la céphalisation de notre lignée on considère que le volume a triplé en 3 Ma, de A. *afarensis* à *Homo sapiens*.

[14] Derrière cette posture intellectuelle de la différence de nature il me semble que l'on doit ranger ceux qui estiment qu'il y a plus de différence entre l'homme et les animaux qu'à l'intérieur de cette même classe animale (des insectes aux mammifères par exemple). Ce serait par exemple le cas d'Heidegger, pour qui l'animal est « pauvre en monde » alors que l'homme lui est « ouvert au monde ». Si l'on comprend largement ces raisons, le problème est que ces derniers sous-entendent et suggèrent un statut quasi hors monde de l'humanité et en fin de compte renouent avec une autre origine de cette dernière, comme si la révolution darwinienne n'avait pas eu lieu. Cela ne change pas au fait de la criante différence anthropologique, il s'agit d'une question d'origine. La question de la différence de degré ou de nature au niveau des comportements culturels me semble toute autre : cf. Dominique LESTEL, *Les origines animales de la culture*, (2003), Champs Flammarion, Manchecourt, 2006, pp. 161-165, pour qui une telle question peut apparaître comme dépourvue de sens.

[15] "Although there was no break in the continuity of universal process of evolution, there was a break in its direction. That is to say, it was not the Ape who continued on in Man, but it was the process of evolution which continued on in its upward movement, and by unfolding a new principle created Man as the representative of a new order of life on earth." P. ALSBERG, *op. cit.*, X, p. 82.

Toute *émancipation somatique* est étrangement refusée à l'Animal[16], permettant alors une division radicale entre l'humanité et le reste du vivant. Ce n'est pas que des pratiques techniques n'existent pas, pour l'auteur, dans le monde animal, mais plutôt qu'elles n'auraient pas et, semble-t-il, ne pourraient pas engendrer les *feedback* que l'on constate chez l'homme. Ces derniers étant, si l'on veut, par nature, essentiellement enfermés dans la prison biologique de l'adaptation somatique :

> "Tool-use, it appears, was here *integrated into the evolutionary scheme of Man*, and it is by virtue of its incorporation in his evolutionary mechanism that Primeval Man's simple act of using stones in defence gains it significance and dynamic momentum: a matter of supreme concern in the vital task of securing his survival, a token of his new evolutionary principle coming into operation, and a promising start to his unique career. With the Animals, on the other hand, it is the principle of body-compulsion, in the sense of body adaptation to environment, upon which evolution converges. From this all-dominant principle the Apes and Monkeys make no exception, however much we may admire their considerable aptitude for using stones as tools and weapons."[xxiv]

Cette façon d'envisager une telle amplification de la distinction Animal / Humanité reste, somme toute, assez fragile. Il suffit de voir comment l'auteur défend une sorte de parti pris : "stone-using Man acts *within* and *for* his principle of evolution; stone-using Ape acts *without* and *against* his principle of evolution. (…) With Man alone has tool-use encroached upon, and was integrated into, the scheme of his evolution, thus driving a wedge between him and the animal world."[xxv] On a l'idée d'une intégration de ce principe, avec ses *feedback*, au sein même de l'humanité, là où le phénomène n'a pas les effets escomptés chez l'animal du fait d'un autre principe évolutif. Cette division, cette distinction semblent donc se faire *par principe*. Dès lors se pose la question suivante : si l'homme a été un animal ; si l'on parle de sa spéciation par la technique ; si ce passage est possible, qu'est-ce qui pourrait bien permettre de l'interdire à d'autres espèces, rangées éternellement et, à vrai dire, arbitrairement dans la prison biologique du *body-adaptation* et du *body-compulsion* ? Il semble que l'auteur admette par pétition de principe un monopole atemporel de l'homme quant au schème évolutif du *body-liberation*. Bien qu'Alsberg soit capable d'être critique quant à la lecture philosophique pré-darwinienne des choses, avec son dualisme corps/esprit[xxvi], il ne témoigne pas moins d'une résistance psychologique, somme toute, très naturelle, à traiter l'homme et l'évolution humaine comme un phénomène *en droit* naturel et normal participant du caractère des espèces "ingénieures" et de l'*ecosystem engineering*.

On s'étonne donc de ce nouveau dualisme modernisé. Au-delà de ces remarques, peut-on attribuer à Paul Alsberg une certaine novation, en tant que précurseur quant à ce qu'on appelle aujourd'hui la "*niche construction*" (Odling-Smee, 1988, 2003)[xxvii] ?

---

[16] "*No Animal makes tools, no Animal speaks in words, or thinks in concepts*. Here is a clear line of demarcation separating Man from the Animal – a dividing line which is safely drawn and only varies with the cultural level of civilization. For if we go backward in human evolution, we will come to a stage when conceptual thought had not yet dawned upon Mankind, as is still the case with certain primitive races of our own times [!!]. (…) Yet, however far back we may trace human evolution we cannot imagine early Man without his artificial tools. As the art of tool-making is unknown to the Animal and Man has always concentrated upon the use of artificial tools, it follows that human life has been distinct from animal life since its inception." (*Ibid.*, III, pp. 16-17).

### 3.2) La question de la construction de niche

La prise en compte, par notre savant, de l'entrée de la technique dans le jeu de l'évolution, insufflant une nouvelle orientation évolutive où la tâche de l'adaptation est désormais transférée à l'outil, ne nous pousse-t-elle pas à reconnaître un *feedback* commun avec la théorie de la *"niche construction"* ? La *niche construction* est un phénomène large : "We describe four major ramifications of niche construction. Niche construction may (1) in part, control the flow of energy and matter through ecosystems (ecosystem engineering), (2) transform selective environments to generate a form of feedback that may have important evolutionary consequences, (3) create an ecological inheritance of modified selection pressures for descendant populations, and, finally (4) provide a second process capable of contributing to the dynamic adaptive match between organisms and environments"[xxviii]. Elle peut être positive, lorsqu'elle augmente la fitness et négative dans le cas contraire. Elle peut être issue d'activités techniques ou « culturelles »[17] comme de simples activités métaboliques. Elle peut ainsi plus ou moins contrer (*counteractive niche construction*[18]) la sélection naturelle classique. Ici, nous sommes, avec Alsberg, dans une partie précise de la *niche construction*, celle de la *cultural niche construction*. Mais, comme l'aspect culturel me semble trop large, on pourrait peut-être préférer parler de *"technaptation"*[19], précisant ainsi le moyen technique (exosomatique) comme facteur principal dans la construction de niche effectuée par certains organismes. Encore, ce terme permet de bien distinguer le fait de construire de celui du feedback qui en résulte, en insistant sur ce dernier, qui est la pièce théorique importante. Il y a alors un phénomène de « niche construction » chez Alsberg lorsqu'il pense l'*émancipation somatique* (*body-liberation*). La surprise est que cet effet rétroactif n'est pensé que pour l'homme. C'est pourquoi, il semble plus juste d'en faire un précurseur de la *cultural niche*

---

[17] "Animal niche construction may depend on learning and other experiential factors, and in humans it may depend on cultural processes." F. J. ODLING-SMEE *et al*, *Niche Construction*, p. 21.

[18] "Conversely, if an environmental is already changing, or has changed, organisms may oppose or cancel out that change, a process we describe as *counteractive* niche construction. More precisely, organisms express counteractive niche construction when they either wholly or partly reverse or neutralize a prior change in an environmental factor." F. J. ODLING-SMEE *et al*, *Niche Construction*, p. 46.

[19] Néologisme construit en 2005 dans mon Mémoire de Master II « Biologie évolutive et théorie de la civilisation : l'effet réversif en question », *Université Lyon 3*, qui m'a semblé utile pour insister sur la composante *"feedback"* résultant de l'activité technique. Je pensais en faire la source de modifications ultérieures, permettant « d'expliquer » certains traits semblant largement dépasser le cadre de l'adaptation stricte (néoténie, céphalisation humaine, mode nidicole secondaire), c'est-à-dire insistant sur le côté hypertélique, un peu d'une manière analogue à A. R. Wallace estimant, par exemple, qu'un cerveau de singe aurait suffit pour la tâche de l'adaptation. Ceci étant, la technaptation doit ce comprendre non pas comme une simple adaptation par moyens techniques, mais comme un processus permettant une acquisition possible de traits dérivés (aptations), de nouvelles « adaptations », permis par la possession d'une technique transformant la structure des pressions de sélection (par baisse des pressions de prédation opérants sur l'espèce en question). Une technaptation voulant nommer la modification des pressions de sélection *par l'utilisation d'outils* **et** *l'ouverture évolutive à d'autres traits* du fait de ce transfert adaptatif (du somatique au technique). En ce sens, il y a chez Alsberg une dynamique évolutive somatique et une dynamique évolutive technaptative, celle de l'humanité. Le moment technaptatif correspondrait chez Alsberg au *délestage somatique*, au fait que le corps suit *secondairement* l'adaptation. À l'adaptation par moyens techniques (exosomatiques), qui est le premier pôle de la *cultural niche construction* (par voie technique), la technaptation veut insister sur le deuxième pôle, le deuxième moment, le feedback et les possibilités qu'elle ouvre pour l'évolution ultérieure de l'espèce techniquement constructrice de sa niche. La théorie de la *niche construction* vient ainsi préciser le côté flou et hésitant de notre conception première. C'est la théorie que je cherchais à construire et c'est pourquoi Alsberg et Odling-Smee me semblent d'importants auteurs.

*construction* (au sens large) que de la *niche construction*, cette dernière étant un phénomène général et généralisé dans le vivant.

Qu'Alsberg cite C. H. Waddington[xxix], – connu en partie, avec S. Wright pour l'imagerie autour du *paysage* en biologie : le paysage adaptatif (*adaptive landscape*, Wright, 1932) et le paysage épigénétique (*epigenetic landscape*, Waddington, 1940)[xxx] – ne doit pas être pris comme un fait important vu qu'Alsberg écrit son livre en 1922 et que l'essentiel y est déjà, à en croire Sloterdijk. Si l'on tient à trouver quelques précurseurs antérieurs dans cette lignée qui mène à la *niche construction* et plus précisément à la *cultural niche construction*, prise dans son pôle « activités techniques », il faut faire référence à A. R. Wallace (1864) et F. Engels (1876). Le premier développe un report des pressions de sélection du corps à l'esprit, résultant d'une adaptation désormais essentiellement technique, là où l'animal ne peut s'adapter que par le corps, et donc bien moins directement. Le second propose en 1883 dans *Dialectique de la nature*, une spéciation de l'homme par l'activité technique, ou, plus précisément, par l'activité de transformation du milieu qu'il entend par *le travail*. Le titre de cette section ne laisse pas subsister le moindre doute : « Le rôle du travail dans la transformation du singe en homme » datant de 1876. Engels y transpose la logique dialectique hégélienne dans la nature – Hegel ne voyait dans la nature qu'une force caduque et condamnée à reproduire le semblable –, aidé par la révolution transformiste et développe ainsi, dès le début de cette section, l'idée que le travail est bien plus qu'une activité de transformation de la matière : « il est infiniment plus encore. Il est la condition fondamentale première de toute vie humaine, et il l'est à un point tel que, dans un certain sens, il nous faut dire : le travail a créé l'homme lui-même. » Engels voit dans cette activité, non plus simplement un rôle essentiel dans la structuration des rapports sociaux, mais bien une force de spéciation, un facteur d'hominisation. Le travail est certes anthropogène, mais cette fois, dans une temporalité évolutive et non plus existentielle ou simplement historique. Pour Engels, refuser le rôle du travail, donc de l'*activité* vitale, aurait un aspect idéologique. Tel est l'enjeu que pointe Odling-Smee soulignant le paradigme de certains psychologues évolutionnistes : "Current use by some evolutionary psychologists of the concept of the environment of evolutionary adaptedness (EEA) is unsatisfactory in another respect. It treats humans as passive victims of selection rather than as virtuoso niche constructors."[xxxi]

3.3) La question biopolitique

Ce nouveau principe d'évolution doit-il amener à considérer invalides la lecture darwinienne de la société et l'ancienne revendication de darwinisme social ? La réponse est affirmative et l'on pourrait même imaginer que le problème moral aurait pu être à l'origine d'une telle théorie. Il est en effet clair que la révision des prémisses rend caduque les conclusions issues de la transgression de la loi de Hume et aboutissant à une volonté de rendre conforme la société humaine à la logique vitale universelle telle que l'a perçue Charles Darwin**.** *Si* l'adaptation humaine ne repose plus strictement sur le corps, mais désormais sur la technique, *alors* le culte de la *lutte pour l'existence* – soit le fait de compter sur les variations et le « fouet » de la sélection pour le bien de l'espèce en évolution – ne tient plus vraiment ou, tout du moins, s'en trouve fortement diminué et à réexaminer. Ne pas s'en rendre compte, ce serait confondre deux plans différents et vouloir recréer les conditions d'une stratégie ancienne et dépassée – celle de l'animal non-humain – avec une espèce qui s'en est émancipée par une stratégie évolutive supérieure. Même si l'auteur n'insiste pas autant sur ce point, il reste possible de faire dériver cette remarque du passage suivant :

> "Sociologists (…) failed to realize that the uniqueness of the human process of evolution, as a natural event, must have been due to an equally unique principle that was at the root of the process. Instead, they contented themselves with the superficial argument that the same factors that were found to be operating in animal evolution must naturally hold good also of human evolution. (…) Such being the state of affairs, with the Darwinian Principle of the "survival of the fittest" figuring as the natural agent of human evolution and with a "Superman" set up as its natural goal, it is hardly surprising that a number of writers and politicians would for the sake of the creation of Superman even recommend a speeding-up of the selectionist process by deliberate extermination of what they think to be the physically or mentally unfit."[xxxii]

La thèse d'Alsberg repose sur l'hypothèse selon laquelle le schème évolutif humain étant fondamentalement différent, les spéculations visant à l'amélioration biologico-évolutive de l'humanité se trompent de principe en exportant le modèle animal dans un domaine où il n'a plus sa validité. Bien entendu, cela est plus efficace et convaincant que de simplement souligner la dite "loi de Hume", estimant le passage du descriptif ou du factuel au prescriptif, envers toutes nos habitudes mentales, nos pratiques et l'histoire, comme inconsistant. Il est en effet plus efficace de trouver l'erreur, non dans le raisonnement – un raisonnement qui d'ailleurs aurait tendance à suggérer que la morale et l'action relèveraient du domaine de la *logique* –, mais dans l'élément descriptif, dans l'*explanans* (l'ensemble des prémisses), dont on voudrait faire une connaissance pour l'action *efficace*.

La venue du surhomme, pour Alsberg, ne saurait se trouver dans l'excellence biologique mais bien plutôt dans le fait de se trouver maître de sa destiné et parfaitement adapté selon des moyens désormais éthiques, rationnels, et techniques là où l'animal n'est adapté que par l'instinct et le somatique.

> "In this final stage of his evolution, in which life is deliberately based on ethical consciousness, Man will be "extra-bodily" adapted to Nature as perfectly as the Animal is "bodily" adapted. Then the great movement which converted Man from an irrational into a rational being, from instinct-ruled life into a free ethical existence, from a slave into a master of his destinies, will have reached, *on an ethical basis*, the ideal equilibrium of adaptive integration that we find with the Animal on an instinctive basis. Then, in terms of biological evolution, the stage of "Superman" is reached."[xxxiii]

La viabilité de l'homme, son « adaptation », ne reposent plus, et de moins en moins, sur le biologique. Toutefois il n'est pas illogique d'estimer, comme Wallace, que la pression de sélection se reporte sur le cognitif, le cérébral, et dès lors penser que le darwinisme (pris comme modèle de lecture du social et de ses pressions de sélection) peut rester un modèle légitime, nécessitant un simple changement de référentiel en relation avec les traits nouvellement adaptatifs[xxxiv], ce qui ne constituerait pas plus une hypothèse *ad hoc*, que celle d'Alsberg[20].

---

[20] Mais quelle est l'hypothèse *ad hoc* ? Est-ce celle d'Alsberg qui permet de maintenir la vision anthropologique du paradigme pré-darwinien ou celle du report des pressions de sélection du biologique au cognitif qui permettant de maintenir, au sein de l'humanité, le paradigme darwinien ? Ces deux thèses réaffirment-elles leurs partis pris envers l'étonnant phénomène humain ? Question de référentiel théorique, ce qui impliquerait la connaissance de la thèse vraie. Malgré cette ignorance, on devrait toutefois choisir la théorie dont l'extension explicative est la plus grande ; celle qui, par voie de conséquence, contient une force

### 3.4) Ce qui manque à Alsberg

Ce qui manque à Alsberg est inhérent à son dualisme Homme/Animal, qui, s'il fait la force de sa théorie et est au centre de son ouvrage, n'en constitue pas moins une rupture peu respectueuse de la phylogenèse et des capacités réelles des êtres vivants. Par exemple, il manque ce que l'on pourrait appeler un *transfert adaptatif du biologique au cognitif*, comme intermédiaire nécessaire au *transfert adaptatif du biologique au technique*. Ce type de transfert est déjà présent chez A. R. Wallace[xxxv], dès 1864, et Charles Darwin y adhère dans son *Descent of Man*[xxxvi]. Il est d'ailleurs possible d'estimer que le transfert adaptatif du biologique au technique comprend le transfert adaptatif du biologique au cognitif, vu que, pour les savants de l'époque, la supériorité technique d'une nation ou d'une tribu sur une autre devait, selon eux, être le témoignage d'une supériorité des facultés mentales des individus composant cette même entité. Cependant, lorsqu'on insiste sur le cognitif et non simplement sur le technique, on évite l'écueil qui consiste à isoler l'homme du monde animal, écueil dans lequel tombera pourtant A. R. Wallace, vers 1866, pour d'autres raisons[xxxvii]. À vrai dire Alsberg cite pourtant A. R. Wallace, mais, ne souligne que sa position sur le pont infranchissable existant entre l'Homme et l'Animal et non sur le fait qu'il s'installe indéniablement dans son prolongement théorique, celui que nous avons choisi d'appeler le « transfert adaptatif ». C'est ce fort clivage entre un *principe de contrainte somatique* et un *principe d'émancipation somatique* qui fait qu'Alsberg considère cette partie technique de la future *niche construction* comme spécifiquement humaine ; clivage accompagné d'un certain finalisme mettant l'homme sur un piédestal, comme quelque chose que l'évolution devait chercher à atteindre.

## Conclusion

Malgré les critiques précédemment faites sur la théorie de P. Alsberg, il y a lieu de mettre en avant la saisie d'une logique évolutive qui aboutit, à mon sens, à l'importante théorie de la *niche construction* qui redonne toute son activité et toute sa force au vivant. En fait, Alsberg réaffirme une vision de "l'étrangeté" humaine qui, pourtant, s'avère intéressante et féconde. Le résultat principal, et étroitement lié à l'emphase sur la différence anthropologique, est ce changement dans les rapports de production humanité/technique. La temporalité existentielle nous démontre que c'est bien l'homme qui produit la technique, même si l'on voit bien l'influence qu'elle a en retour sur lui. La temporalité évolutive, par contre, pourrait témoigner du fait "contre-intuitif" d'une technique capable de "produire", de rendre possible l'humanité avec ses traits distinctifs, comme sa "nudité" (au sens large), sa néoténie[xxxviii] et sa céphalisation rapide. Ainsi, l'humanité ne ferait pas que produire des techniques, elle ne serait pas qu'*Homo faber*. Apparaissant alors comme une humanité *technosymbiotique*[xxxix], c'est-à-dire produite par la nouvelle dynamique qu'implique la possession d'une technique. Ce renversement des rapports de production constitue une avancée philosophique intéressante et

---

d'unification supérieure, une unité plus vaste, segmentant moins le réel que sa rivale. Or, la seconde thèse permet de comprendre l'activité vitale et les traits humains, alors que la première scinde le monde et l'évolution pour en rendre compte. Quant à la théorie de la *construction de niche,* elle a une valeur supérieure, en montrant que l'activité des espèces cause ce report des pressions de sélection et l'*apparent* détachement de certaines espèces « constructrices de niches » par rapport à ce qu'on pourrait attendre du fait de son biotope. Cet écart pourrait ainsi être la *signature* de la *niche construction*.

profonde, mettant en valeur une technique qui ne serait plus désormais une simple production culturelle, mais un véritable facteur d'hominisation[xl]. Tout cela s'inscrit et s'apprécie pleinement dans la logique contemporaine de la *construction de niche*. Peut-être aurons-nous alors convaincu le lecteur qu'il y a bien un phénomène de « niche construction » chez Alsberg lorsqu'il pense l'*émancipation somatique*.

La recherche de précurseurs est toujours chose délicate, car il faut choisir certaines pièces théoriques plutôt que d'autres. A. R. Wallace, comme F. Engels, ont leur place dans ce fil d'Ariane reliant la biologie évolutive de Darwin à Odling-Smee. Toutefois, les deux premiers auteurs, au même titre qu'Alsberg, pourraient être autant acceptés qu'écartés, ne pensant cette rétroaction que pour l'homme, alors que même la *cultural niche construction* est plus ouverte. Quoiqu'il en soit, c'est une chose de penser tels ou tels concepts explicatifs, d'en faire le centre de la théorie ou un usage anecdotique, mais cela en est une autre de fournir les exemples, le programme de recherche et de vérification de la théorie ; d'exposer la force explicative et les anomalies que cette dernière résout, dans une pure démarche de construction scientifique. Malgré tout, peut-être qu'Odling-Smee *et al*, verront, si ce n'est déjà fait, en Paul Alsberg un de leur ancêtre par une parenté théorique quant au "*neglected process*" de l'évolution.

<div style="text-align:center">

\*

\* \*

</div>

---

[i] Voici ce qu'en dit Peter Sloterdijk : « C'est Paul Alsberg qui, avec son livre *L'énigme de l'humanité*, en 1922, a posé la pierre essentielle d'une théorie du devenir-humain, même si les spécialistes de ce secteur ont fait de ce livre un usage réduit. Dans ce qu'il appelle la suspension des corps [traduction de "*body-liberation*" et en allemand de « *Körperausschaltung* »], il a discerné le mécanisme clef de l'anthropogenèse. Il s'agit ici d'un concept avec lequel l'histoire naturelle passe de la prise de distance avec les environnements naturels, dans la lignée des insulations spontanées, à une première histoire de la distanciation à l'égard de la nature, sur la lignée d'un usage de l'outil d'abord spontané, puis élaboré et chronique. Le théorème d'Alsberg interprète le devenir-humain comme l'effet d'une hyper-insulation, dont l'effet central consistait à émanciper le pré-homme de la nécessité de s'adapter du mieux qu'il le pouvait à son environnement. On a, à juste titre, qualifié l'événement décrit par la suspension des corps comme la **« sortie de la prison »** [note de l'auteur : *Cf.* Le titre de la nouvelle édition allemande : *Das Menscheiträtsel : Der Ausbruch aus dem Gefängnis – Zu den Entstehungsbedingungen des Menschen*, avant-propos de Dieter Claessens, Giessen, 1975] formée par la relation biologiquement déterminée avec l'environnement. » P. SLOTERDIJK, *La Domestication de l'Être*, Mille et une nuit, 2000, p. 48. Même si Yves Michaud ne parle pas directement d'Alsberg (MICHAUD, 2006, pp. 46-47), on consultera tout de même son ouvrage *Humain, inhumain, trop humain*, Climats (Flammarion), Paris, 2006, qui permet de s'imprégner de la philosophie de Sloterdijk faisant de la technique un élément central et tirant quelques implications de la philosophie de la technique ici développée.

distinction entre les causes proches et lointaines, les causes mineures et majeures, causes externes (géotope, biotope, sociotope) et internes (génétique, "contraintes architecturales", allométrie). Il ne faudrait pas non plus oublier les apports décisifs de la paléoanthropologie et prendre toute la mesure de la découverte de l'équipe de Michel Brunet au TCHAD, le 19 Juillet 2001, par Ahounta Djimdoumalbaye, du crâne de Toumaï (env. 7 Ma). On se référera aux articles de Michel Brunet publiés dans la prestigieuse revue *Nature*, et notamment, M. BRUNET, F. GUY, D. PILBEAM, *et al.*, "A new hominid from the Upper Miocene of Chad. Central Africa," *Nature*, **418**, 145-151 (2002). En langue française, on pourra se reporter au chapitre V « Le soleil de l'humanité se lève aussi à l'Ouest », du livre de M. Brunet, *D'Abel à Toumaï*, Odile Jacob, Mayenne, 2006.

xviii "power of shaping his environment" ALSBERG, *op. cit*., XIX, p. 170.

xix "Man (…) stands outside and above the Animal, but inside and below Nature." (*Ibid.*, XIX, p. 170).

xx « notre cerveau, notre société et notre langage ne sont que les signes extérieurs et divers d'une seule et même supériorité interne. Ils disent, chacun à sa manière, le succès unique, exceptionnel, que la vie a remporté à un moment donné de son évolution. Ils traduisent la différence de nature, et non pas seulement de degré, qui sépare l'homme du reste de l'animalité. Ils nous laissent deviner que si, au bout du large tremplin sur lequel la vie avait pris son élan, tous les autres sont descendus, trouvant la corde tendue trop haute, l'homme seul a sauté l'obstacle. C'est dans ce sens tout spécial que l'homme est le « terme» et le « but » de l'évolution. » Henri BERGSON, *op.cit.*, p. 265. Ou encore, peu après ce dernier passage : « L'ensemble du monde organisé devient comme l'humus sur lequel devait pousser ou l'homme lui-même ou un être qui, moralement, lui ressemblât. » (*Ibid.*, p. 267).

xxi ALSBERG, *op. cit.*, XXI, p. 199.

xxii *Ibid.*, XIX, p. 167.

xxiii *Ibid.*, XXI, p. 196.

xxiv *Ibid.*, X, pp. 77-78.

xxv *Ibid.*, X, p. 81.

xxvi Sur ce point, voici le sentiment de P. Alsberg : "The philosophical axiom that Mind belongs to another category than the body, because the body is "extended in space" and Mind is not, disregards their integral biological unity. Function is essentially a Life-manifestation, an *organ in action*, as it were, and as such existing in, and with, the organ, and not somewhere else. To say that the eyes are "extended" and seeing is not, or that the legs are extended in space and walking is not, reveals stark unbiological thinking." (*Ibid.*, XXI, p. 206). Ce qui ne l'empêche d'ajouter ensuite, sur la même page et contre toute attente : "Its collapse does not, however, necessarily imply the total invalidity of the old dogma of the heterogeneity and immortality of the human "soul"".

xxvii F. John Odling-Smee note, comme suit, les auteurs importants dans la construction et l'appui de sa théorie : E. Schrödinger (*Mind and Matter,* 1944) ; E. Mayr (*Animal Species and Evolution,* 1963) ; Conrad Waddington (1959, 1969) ; Gould and Lewontin (*The Spandrels of San Marco and the Panglossian Paradigm : A critique of the Adaptationist Programme*, 1979) ; R. Dawkins (*The Extended Phenotype*, 1982) ; puis Levins and Lewontin 1985 ; Wilson 1985 ; West-Eberhart 1987 ; West et al. 1988 ; Bateson 1988 ; Plotkin 1988 ; Wcislo 1989 ; Hot and Gaines 1992 ; Michel and Moore 1995; Brandon and Antonovics 1996 ; Moore et al. 1997 ; Wolf et al. 1998 ; Oyama et al. 2001 ; Sterelny 2001 ; Griffiths and Gray 2001.

xxviii F. J. ODLING-SMEE *et al*, *Niche Construction*, pp. 2-3.

xxix "Animals, he urged [C. H. WADDINGTON, *The Nature of Life*, 1961], were well capable of selecting *for themselves* their particular environment and thereby would gain a decisive influence on the type of natural selective pressure imposed upon them by the new environment. In his theory *animal behaviour* in its relation to environment is thus taken to have effectively contributed to the course taken by evolution, which in a way is plain Lamarckism, yet in the modern sense that the animal behavior is conditioned by hereditary factors enwrapped in the chromosomes. The basic question here is how far is animal behaviour, in response to environment, influenced by hereditary constitution, and how far may it, on its part, **itself influence the direction of mutational changes?** The question is not readily accessible to experimental investigation; but the possibility of *directed mutation* is very much in the mind of modern geneticists, and some promising experiments have already been carried out in this line." P. ALSBERG, *op. cit.*, XXI, p. 190. Il est d'ailleurs à noter que c'est Waddington lui-même qui rédige la courte préface de son ouvrage.

xxx Cf., par exemple, Jean GAYON, « La marginalisation de la forme dans la biologie de l'évolution », *Bulletin d'histoire et d'épistémologie des sciences de la vie*, vol. 5, n°2, 1998, pp. 133-166. « Dans l'histoire contemporaine des sciences, le paysage a souvent été une image de ce que l'on appelle classiquement un système dissipatif, c'est-à-dire un système dynamique dans lequel un mouvement quelconque s'achemine vers une position d'équilibre. Quelles que soient les conditions initiales, le système se retrouvera au voisinage d'un équilibre au bout d'un certain temps. (…) Aussi le potentiel d'un système dissipatif peut-il être représenté comme un "paysage" ou une "topographie" dans laquelle des "bassins" représentent des équilibres stables, et sont séparés par des montagnes et des "cols", qui figurent des équilibres instables (…). Il convient de noter que Wright inverse la polarité classique de l'image : au lieu de descendre vers des bassins, les populations mendéliennes montent vers des "pics adaptatifs". » (*Ibid.*, p. 145).

xxxi F. J. ODLING-SMEE *et al*, *Niche Construction*, p. 368.

xxxii ALSBERG, *op. cit.*, XX, p. 174.

xxxiii *Ibid.,* XX, p. 180.

xxxiv Cf. l'article de Philippe HUNEMAN, « Les difficultés du concept d'adaptation », *Bulletin d'histoire et d'épistémologie des sciences de la vie*, vol. 12, n°1, 2005, pp. 173-199.

xxxv "From the time, therefore, when the social and sympathetic feelings came into active operation, and the intellectual and moral faculties became fairly developed, man would cease to be influenced by "natural selection" in his physical form and structure ; as an animal he would remain almost stationary ; the changes of the surrounding universe would cease to have upon him that powerful modifying effect which it exercises over other parts of the organic world. But from the moment that his body became stationary, his mind would become subject to those very influences from which his body had escaped ; every slight variation in his mental and moral nature which should enable him better to guard against adverse circumstances" Alfred R. WALLACE, "The Origin of Human Races and the Antiquity of Man deduced from the theory of "Natural Selection."", *Journal of the Anthropological Society of London*, Vol. 2. (1864), pp. clxiii-clxiv. Au paragraphe suivant l'idée de transfert adaptatif est exprimée clairement : "But from the time when this mental and moral advance commenced, and man's physical character became fixed and immutable, a new series of causes would come into action, and take part in his mental growth. (…) When the power that had hitherto modified the body, **transferred its action to the mind**, then races would advance and become improved merely by the harsh discipline of a sterile soil and inclement seasons." Alfred R. WALLACE, *op. cit.*, p. clxiv. Il est intéressant de faire remarquer que, d'une part, Wallace note à la fin de l'article une dette envers Spencer et, d'autre part, que H. Spencer prend en compte son article dans ses *Principes de biologie*, Tome I, (1864), Félix Alcan, Paris, 1893, § 170, note 1, (fin du chap. XIII), p. 568. Ce dernier défend l'expansion de l'équilibration directe contre l'équilibration indirecte, que l'on pourrait traduire en notre langage comme le transfert d'une adaptation passive, indirecte (dont la sélection naturelle), à une adaptation active, reposant sur l'hérédité de l'acquis, certes, mais principalement sur le système nerveux (*Principes de biologie*, t. I, p. 567). Il semble donc que l'idée de transfert adaptif du « biologique » au cognitif, leur appartienne à tous les deux.

xxxvi Voir Charles DARWIN, *The Descent of Man*, (1871, 1874), début du *chap*. V, "On the development of the intellectual and moral faculties during primeval and civilized times".

xxxvii Cf. Robert J. RICHARDS, "Darwin on Mind, Morals, and Emotions", *The Cambridge Companion to Darwin*, eds. J. Hodge and G. Radick (Cambridge: Cambridge University Press, 2003), pp. 92-115. (Sur le site de l'auteur, pp. 1-34). "Wallace wrote Darwin (18 April 1869) to say that his altered view about human evolution derived from his empirical testing of medium's power" (note 39, p. 33). R. J. Richards nous indique que c'est "a superior intelligence" qui, pour Wallace, "has guided the development of man in a definite direction, and for a special purpose, just as man guides the development of many animal and vegetable forms." (Wallace, 1891a, 204). La perception, chez Wallace, de l'hypertélie (on parle d'hypertélie mais aussi d'hypermorphose (syn.) lorsque on estime qu'un trait dépasse largement son but adaptatif) du cerveau du « sauvage » est aussi à prendre en compte.

xxxviii Cf. Konrad LORENZ, « Psychologie et phylogenèse » (1954), *Trois essais sur le comportement animal et humain*, édition du seuil, Manchecourt, 1974, en particulier pp. 223-232.

xxxix En général, cette technosymbiose est projetée dans un avenir relativement proche, comme quelque chose de plus ou moins inévitable, suscitant divers types de craintes et d'espoirs lorsqu'elle n'est pas pensée comme simple fantasme, en dépit des énormes possibilités de la "technoscience" (G. Hottois) contemporaine. L'intérêt dont témoignent les philosophes envers les biotechnologies et la notion d'*anthropotechnie*, atteste l'enjeu et la modification de nos représentations touchant à la vie humaine, à l'éthique de la médecine et foule d'autres considérations fondamentales (Cf. P. Sloterdijk, *La Domestication de l'Être* ; Jérôme Goffette, *Naissance de l'anthropotechnie, de la médecine au modelage de l'humain* ; Yves MICHAUD *Humain, inhumain, trop humain* ; Jean-Michel TRUONG, dans son *Totalement inhumaine* et ailleurs, qui n'hésite pas à développer cette image provocante et inquiétante du développement actif de notre "Successeur". L'idée ici défendue est que l'anthropotechnie contemporaine, avec son accélération, ses probables dérives et l'engrenage dans lequel elle emporte l'humanité de ce millénaire, n'est que la partie émergée de l'iceberg. Ce qui signifie, dans le langage de Sloterdijk, (tel que le décrit GOFFETTE, *op. cit.* p. 25), que l'*anthropotechnie secondaire* (volontaire), n'est que la partie émergée de l'*anthropotechnie primaire*, celle qui comprend l'homme comme produit et permis par la technique et les insulations successives vis-à-vis de l'environnement qu'elle rend possible.

xl Cette idée est, me semble-t-il, loin d'être rejetée par Odling-Smee : "For most vertebrates, the role of acquired characteristics in evolution is likely to be fairly restricted, but there is every reason to believe that acquired characters may have been important in hominid evolution. The models that we described in chapter 6 revealed circumstances under which cultural transmission could overwhelm natural selection, accelerate the rate at which a favored allele spreads, produce novel evolutionary events, and perhaps have triggered hominid speciation. We argue that, because cultural processes typically operate faster than natural selection, cultural niche construction is likely to have more profound consequences than gene-based niche construction." F. J. ODLING-SMEE *et al, Niche Construction*, p. 377.

*P.S.* Les remarques et éventuelles corrections sont toujours les bienvenues. Merci par avance.